# Superrotation on Venus: Driven By Waves Generated By Dissipation of the Transterminator Flow

Héctor Javier Durand-Manterola

Departamento de Ciencias Espaciales
Instituto de Geofísica
Universidad Nacional Autónoma de México

hdurand@geofisica.unam.mx

**Abstract**

*Context:* The superrotation phenomenon in the atmosphere on Venus has been known since the late 60's. But until now no mechanism proposed has satisfactorily explained this phenomenon.

*Objective:* The aim of this research is to propose a mechanism, until now never considered, which could drive the atmosphere of Venus in its superrotation. This mechanism involves the transfer of the transterminator ionospheric flow momentum to the lower atmosphere via pressure waves generated in the cryosphere of Venus. The mechanism proposed presents a source of energy sufficiently strong to allow the transfer of energy despite dissipation.

*Method:* The energy flow which transports the transterminator flow and the energy lost by the viscosity in the superrotating atmosphere were calculated. Both results were compared to establish if there is sufficient energy in the transterminator flow to drive the superrotation. Finally, the amplitude that the waves should have to be able to obtain the momentum necessary to induce superrotation was calculated. Also an experimental model was made presenting some similarities with the process described.

*Results:* The calculated power for the transterminator flow is ~$8.48 \times 10^{10}$ W. The calculated viscous dissipation of the superrotating flow is ~ $1.4 \times 10^9$ W. Therefore, there is sufficient energy in the transterminator flow to maintain superrotation. The amplitude of the waves generated in the cryosphere, necessary to deposit the power dissipated by the viscous forces, is $10^{-4}$ m for waves of 1 Hz and $10^{-8}$ m for waves of $10^4$ Hz. These amplitudes imply that at the altitude of the clouds on the night side there must be a constant sound of 83 dB.

If the superrotation of Venus were to stop, with the continuous injection of $1.4 \times 10^9$ W, the actual superrotation would appear again in $1.4 \times 10^6$ years.

# 1 Introduction

The atmosphere of Venus, at the altitude of the clouds, takes 4 days to go all the way round (Gierasch et al., 1997), whilst the solid body of the planet takes 243.0187 days (Bakich, 2000). This phenomenon is known as atmospheric superrotation. The Venus superrotation phenomenon has been known since the 60's (Young and Schubert, 1973) but until now (April 2010) no one has been able to find a viable mechanism which can explain it. Several mechanisms have been proposed, amongst them, tidal waves (Gold and Soter, 1971), high latitude jet streams (Del Genio and Shou, 1996), differential warming (Yamamoto and Takahashi, 2003) etc. but none of these has been completely satisfactory.

Titan is another body which also presents superrotation in its atmosphere (Bird et al., 2005; Del Genio and Shou, 1996; Hourdin *et al.* 1995; Del Genio *et al.* 1993). It is thought that maybe the mechanisms responsible for the superrotation of both planets are similar.

In this work it is suggested that superrotation on Venus is driven by the waves generated in the cryosphere when the transterminator flow dissipates. The interaction between the flow on the dawn side and the flow on the dusk side generates waves. These flows have different speeds, that of the dusk side being much quicker, therefore, by momentum conservation, the waves in the retrograde direction have a greater momentum than in the prograde direction. These waves reach as far as the troposphere, deposit their momentum at this level and drive the superrotation.

The energy obtained from the transterminator flow was calculated as two orders of magnitude greater than the energy dissipated by the viscous forces in the superrotating flow. The amplitude that the waves should have to maintain the superrotation in a stationary state was calculated as values smaller or equal to $10^{-4}$ m. Also calculated was the time it would take the atmosphere to acquire the actual superrotation, starting from co-rotating with the planet.

## 2 Transterminator flow

On Venus exist two flows that travel from the day side to the night side. 1) One cell is in the high mesosphere and in the thermosphere and driven by solar heat. This cell has a relatively stable circulation with a speed of ~ 200 m/s (Bougher et al. 2006), and 2) the so called transterminator flow which is the movement of the ions of the ionosphere at supersonic speed (~ 2 to 4 km/s at a height of 250 km) (Fox and Kliore, 1997). The transterminator flow was discovered in the observations of the Pioneer Venus Orbiter in its first two years of orbiting Venus (Knudsen et al., 1981).

The first flow is a movement from the neutral atmosphere. This is driven from the sub solar point to the antisolar point by the difference in pressure due to the difference in temperature between the day side (~ 300 K) and the night side (~ 100 K). This produces a flow with a velocity of 200 m/s (Bougher et al. 2006).

On the other hand in the transterminator flow in the ionosphere (at an altitude of 150-800km in the terminator) (Miller and Whitten, 1991; Fox and Kliore, 1997) is the ionized component which moves at supersonic speeds (~2 to 4 km/s at an altitude of 250 km) (Fox and Kliore, 1997). Several authors (Whitten et al. 1984; Elphic et al. 1984; Nagy et al., 1991) explain this flow as also driven by the difference in pressure due to the difference in temperature. But Pérez-de-Tejada (1986) demonstrated that a De Laval nozzle is needed, as in the case of the solar wind (Dessler, 1967), so that a difference in pressure can produce a supersonic flow. Given that in the atmosphere of Venus such a nozzle does not exist, then it is necessary to find another explanation for the origin of the supersonic transterminator flow. Pérez-de-Tejada (1986) demonstrated that the transterminator flow momentum is the same as the momentum that the solar wind loses in areas surrounding Venus. Hence, Pérez-de-Tejada (1986) proposed that the supersonic transterminator flow of the ionized material is generated by the viscous dragging of the solar wind in the ionosphere of Venus.

Whichever mechanism drives the transterminator flow, it is a fact that, between 120° and 150° of the sub solar point and at an altitude between 300 and 800 km, the flow on the dusk side is much quicker than on the dawn side, with a difference of up to 2 km/s (Miller and Whitten, 1991). Both flows come in contact with one another on the night side. The interaction between these two flows generates turbulence and waves. As both flows are supersonic it is thought that their interaction generates two shock waves which stop each one flows and dissipate the greater part of their energy in the form of heat. The existence of these two shock fronts has been proposed but it has never been observed. However, it is known that further away than 150° from the sub solar point the transterminator disappears and the flow becomes chaotic (Fox and Kliore, 1997) indicating that this is where the area of turbulence for shocked material is located. Due to the asymmetry in the speed, there is more momentum in a retrograde direction than in a prograde direction, thus when both flows interact more waves are generated in a retrograde direction than in the prograde direction. These waves travel from the ionosphere, through the thermosphere and the mesosphere towards the troposphere depositing momentum en route and dissipating in the cloud layer, depositing here most of the momentum and moving the atmosphere in a retrograde direction.

To see if the model is viable, the power of the transterminator flow must be calculated, and also the power dissipated by the superrotating atmosphere, and then the two compared to see if the power of the transterminator flow is greater than the dissipated power of the superrotating atmosphere. Finally we calculate the amplitude and the intensity in decibels of the necessary waves to replace the dissipated energy in the superrotating flow.

## 3 Analogue experiment

An experiment which illustrates the mechanism proposed here is the following (Figure 1): On a flat sheet made of any material (in this experiment, an expanded polystyrene sheet) a jet of water was allowed to fall from a certain height (0.2 m) which, on reaching the sheet, created a flow which radiated from the jet (2 m/s).

On the other side of the sheet another jet of water was allowed to fall, but from a lower height (0.02 m). This second jet also created a radial flow on the sheet (0.63 m/s),

The interaction between both flows on the sheet created an area of turbulence in which superficial waves could be seen moving from the faster flow towards the slower one. In the slower radial flow waves also moved but not in the fast one, demonstrating that the momentum of the waves goes in the direction previously mentioned. In this experimental model the velocity of the superficial waves is 0.35 m/s, in other words, both jets move at a greater speed than the waves in the middle which is an analogue to the transterminator supersonic flow. This experiment illustrates, although it does not proves, the proposed mechanism.

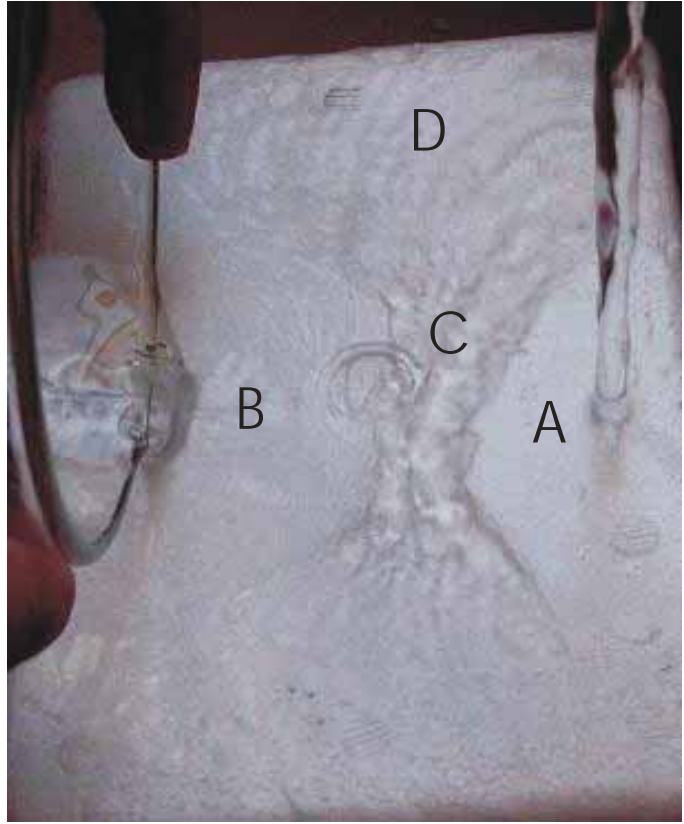

Figura 1
Analog experiment: In the interaction of two water flows (A y B) they form a turbulence region (C) in which it is generated surface waves (D) that travel from the faster flow (A) to the slower one (B). (Image: Irma Yolanda Durand Manterola)

## 4 Transterminator flow energy

The transterminator flow is accelerated, increasing speed from the sub solar point up to the point at which shock waves are generated. The greatest velocity should be attained just before the shock waves are generated, hence the energy per unit area and per unit transterminator flow time on the dawn side $I_{da}$ is

$$I_{da} = E_{da} v_{da} = \frac{1}{2} \rho_{da} v_{da}^3 \qquad (1)$$

Where E is the energy per unit volume, $\rho$ is the density, and v is the velocity. The subindex, da, indicates variables on the dawn side.

On the other hand, the energy per unit area and per unit time for the transterminator flow of the shock wave on the dusk side $I_{du}$ is

$$I_{du} = E_{du}v_{du} = \frac{1}{2}\rho_{du}v_{du}^3 \qquad (2)$$

The du subindex indicates variables on the dusk side.
The power of the flow in a layer at a height h and thickness dh would be

$$dP = \pi(R+h)I_{da}dh + \pi(R+h)I_{du}dh = \pi(R+h)\left(\frac{1}{2}\rho_{da}v_{da}^3 + \frac{1}{2}\rho_{du}v_{du}^3\right)dh \qquad (3)$$

Where R is the radius of the solid body of Venus.
The total power is

$$P = \int_{150}^{800} \pi(R+h)\left(\frac{1}{2}\rho_{da}v_{da}^3 + \frac{1}{2}\rho_{du}v_{du}^3\right)dh \qquad (4)$$

The 150 and 800 values in kilometers are the altitude limits between which the transterminator flow moves (Knudsen and Miller, 1992).
The integral (4) was resolved using speed and numerical density values given by Knudsen and Miller (1992, Figure 1). The velocity values were multiplied by 4 to obtain $v_{du}$ and by 2 to obtain $v_{da}$, because the Knudsen and Miller (1992) values are median values and what is needed for the calculation of power are the extreme values before the shock wave and it is known that these extreme values are 2 and 4 km/s at 250 km whereas the value given by Knudsen and Miller (1992) is 1km/s. When calculating P the result is $8.48 \times 10^{10}$ W. This is the power of the transterminator flow immediately before entering the shock wave.
When both dawn and dusk flows interact on the night side their kinetic energy is transformed into dissipated heat and energy which is transmitted as ion acoustic waves up in the thermosphere and pressure waves lower down in the cloud layer.
As the transterminator flow momentum on the dusk side is greater than on the dawn side, since $v_{du} > v_{da}$, then by momentum conservation most of the waves which are generated will move in a retrograde direction and when they dissipate in the lower atmosphere they will drive this also in a retrograde direction.

The neutral particles, in the cryosphere, do not contribute to the transterminator flow energy because of the following: as the radius of the O atom is $1.46 \times 10^{-10}$ m (Kittel, 1976, p 134), then the transversal section of interaction is $\sigma = \pi r^2 = 6.7 \times 10^{-20}$ m$^2$. On the day side at 250km the density of the particles is of the order $n = 2 \times 10^{10}$ m$^{-3}$ (Kasprasak et al, 1997, p227). Therefore the mean free path $\lambda = \dfrac{1}{n\sigma}$ is $\lambda = 7.46 \times 10^5$ km. The mean free path of the ions is 50 km (Pérez de Tejada et al., 2010) therefore the ion-ion interaction is collisional and the ion-neutral interaction is not.

## 5 Power dissipated by the atmosphere of Venus

The viscous dissipation power $P_d$ is (Shu, 1992, p 51; Fox and McDonald, 1985, p 29)

$$P_d = \mu \left(\frac{du}{dy}\right)^2 \tag{5}$$

This has W m$^{-3}$ units.

In a layer of thickness dy and surface S(y) the viscous dissipation will be

$$dP = P_d S(y) dy = \mu \left(\frac{du}{dy}\right)^2 S(y) dy \tag{6}$$

The superrotating flow region has the shape of a barrel and therefore its surface will be

$$S = 4\pi (R+y)^2 \operatorname{sen}(\lambda) \tag{7}$$

Where R is the planet radius, y is the layer altitude and $\lambda$ is the maximum altitude the superrotating flow reaches ~ 70°.

Substituting (7) in (6)

$$dP = 4\pi P_d (R+y)^2 \operatorname{sen}(\lambda) dy = 4\pi \mu \left(\frac{du}{dy}\right)^2 (R+y)^2 \operatorname{sen}(\lambda) dy \tag{8}$$

And integrating in all altitudes, gives

$$P = 4\pi sen(\lambda)\int_0^H P_d(R+y)^2 dy = 4\pi sen(\lambda)\mu\int_0^H \left(\frac{du}{dy}\right)^2 (R+y)^2 dy \qquad (9)$$

Inserting the values of the speed profile of Venus we obtain P = 1.4x10$^9$ W which is the dissipated power by the atmosphere in superrotation.

It may be seen that the transterminator flow power is greater than the power dissipated by the superrotating atmosphere. Therefore the transterminator flow has sufficient energy to replace the superrotation losses.

## 6 Wave amplitude

When both transterminator flows (dusk and dawn) interact they form an area of turbulence in which ion-acoustic waves are formed. These descend to the cloud layer becoming acoustic waves, and deposit their momentum. Since the dusk flow has more momentum than the dawn flow then, by conservation of momentum, the waves with retrograde momentum will have more power than the waves with prograde momentum driving the superrotation.

The average power flow of the acoustic waves is (Elmore and Heald; 1985; p 144)

$$I = \frac{1}{2}\rho_0 c\left(\frac{\partial \xi}{\partial t}\right)^2 = \frac{1}{2}\rho_0 c\omega^2 \xi_m^2 \qquad (10)$$

Where c is the velocity of sound, $\omega$ is the angular frequency of the waves and $\xi_m$ is the amplitude of the waves.

Calculating the amplitude $\xi_m$ from the equation (10) gives

$$\xi_m = \sqrt{\frac{2I}{\rho_0 c\omega^2}} \qquad (11)$$

The velocity of sound in gases is (Elmore and Heald, 1985)

$$c = \sqrt{\frac{\gamma RT}{M}} \qquad (12)$$

Where $\gamma$ (=1.36) is the adiabatic dilatation coefficient for $CO_2$ at 100 K, (the temperature of the cryosphere on Venus) (Kasprzak et al., 1997), R (= 8.314 J Mol$^{-1}$ K$^{-1}$) is the universal gas constant, T (= 100 K) is the temperature of the cryosphere and M (= 0.044 kg/mol) $CO_2$ molar mass. Therefore the speed of sound in the cryosphere of Venus is 160.3 m s$^{-1}$.

Substituting (12) in (11) gives

$$\xi_m = \sqrt{\frac{2I}{\rho_0 \sqrt{\frac{\gamma RT}{M}} \omega^2}} \qquad (13)$$

Finally, $I = P/S_o$, i.e., the dissipated power between the shock waves surfaces. Therefore (13) becomes

$$\xi_m = \sqrt{\frac{2P/S_o}{\rho_0 \sqrt{\frac{\gamma RT}{M}} \omega^2}} \qquad (14)$$

Inserting values referring to the cryosphere and the cloud layer of Venus, the amplitude values of the sound waves, supposing a power equal to the dissipation P, are $10^{-4}$ m for waves of 1 Hz and $10^{-8}$ m for waves of $10^4$ Hz, showing that the amplitude is sufficient to deposit a power of $1.4 \times 10^9$ W at the altitude of the clouds. Pretty large amplitudes are not needed to deposit this energy.

The structure for three maxima in the velocity profile (Gierasch et al., 1997) supports the idea of the drive of the atmosphere by waves, since any viscous transmission momentum mechanism would produce a profile monotonously growing with altitude. On the other hand, if several frequencies are produced in the cryosphere, and as the absorption coefficient depends on the frequency, then some frequencies descend more than others before dissipating and deposit their momentum at different altitudes. For this reason, the structure for three maxima of the speed of the superrotating flow profile may be explained.

On Titan the zonal winds also present a structure of several maxima, but not a monotonous growing profile (Bird et al., 2005) which leads to thinking that there it may also possibly be driven by waves. But it is not known if in Titan a transterminator flow equivalent exists.

## 7 Wave intensity in decibels

If the power transported by the waves ($1.4 \times 10^9$ W) is divided between the vertical area where the waves pass by (S = $6 \times 10^{12}$ m$^2$), then the intensity of the waves would be $1.92 \times 10^3$ W/m$^2$. If the hearing limit ($1 \times 10^{-12}$ W/m$^2$) is taken as the null level then the intensity level of the waves will be K = 84 dB, calculated via the following equation (Kitaigorodski, 1975)

$$K = 10 \log\left(\frac{I}{I_0}\right) \qquad (15)$$

where I is the intensity of the waves and $I_0$ is the null level.
84dB is a level of intensity of sound similar to that produced by a symphony orchestra playing "fortissimo".

## 8 Superrotation acceleration time

Kinetic energy of the atmosphere in superrotation is of the order of $6.2 \times 10^{22}$ J. So, if starting from a superrotation of zero and the waves deposit $1.4 \times 10^9$ W, then the atmosphere would take $1.4 \times 10^6$ years to reach its actual speed, which on a geological scale is very fast.

## 9 Conclusions

The conclusions obtained from this research are:
The power that the transterminator flow has before the shock wave is in the order of $8.48 \times 10^{10}$ W.
The power which the atmosphere dissipates in superrotation is of the order of $1.4 \times 10^9$ W.
Therefore the energy that the transterminator flow has is sufficient to compensate for the losses from the atmosphere in superrotation.
The waves produced in the area of turbulence which travel in a retrograde direction present a greater intensity than the ones traveling in a prograde direction. This is due to the asymmetry in the velocity between the dawn side and the dusk side and the conservation of momentum.
Therefore when the waves dissipate in the cloud layer the movement they generate in this layer will be retrograde, as is the movement observed.
The intensity of the waves is 84 dB which is sufficient for Venus to maintain in the night zone, at the altitude of the clouds, a roar similar to a symphony orchestra playing "fortíssimo". Maybe in the future a spaceship will be equipped with a microphone to prove this prediction for the model.
Supposing that the atmosphere of Venus rotated in unison with the solid body of the planet and the waves deposited $1.4 \times 10^9$ W, then after $1.4 \times 10^6$ years the atmosphere would superrotate at the actual speed.

Acknowledgements: I thank Deni-Tanibe Zenteno-Gomez and Hector Perez-de-Tejada our useful discussions which led to the ideas expressed in this article. Although they do not want to appear as authors I always consider co-authors of this paper.